\documentclass[reprint,superscriptaddress,amsmath,amssymb,prl]{revtex4-2}
\usepackage{xspace}
\usepackage{bm,physics}
\usepackage{graphicx}
\usepackage{xcolor}
\usepackage{amsmath,amssymb,amsfonts,amsthm}
\usepackage[colorlinks=true,linkcolor=blue,anchorcolor=red,citecolor=blue,urlcolor=blue]{hyperref}
\usepackage{multirow}
\usepackage{diagbox}

\def \K {{\mathcal{K}}}
\def \I {\hat{I}}

\def \H {\mathcal{H}}

\begin{document}
	\title{Realization of Arbitrary Gauge Fields via Symmetry-Protected Zero Modes}
	
	\author{J. X. Dai}
	\thanks{These authors contributed equally to this work.}
	\affiliation{Department of Physics and HK Institute of Quantum Science \& Technology, The University of Hong Kong, Pokfulam Road, Hong Kong, China}
	
	\author{Bingbing Wang}
	\thanks{These authors contributed equally to this work.}
	\affiliation{Department of Physics, The Chinese University of Hong Kong, Shatin, Hong Kong SAR, China}
	
	\author{Jiangzi Chen}
	\affiliation{Department of Physics, The Chinese University of Hong Kong, Shatin, Hong Kong SAR, China}
	
	\author{Y. X. Zhao}
	\email[]{yuxinphy@hku.hk}
	\affiliation{Department of Physics and HK Institute of Quantum Science \& Technology, The University of Hong Kong, Pokfulam Road, Hong Kong, China}
	
	\author{Haoran Xue}
	\email{haoranxue@cuhk.edu.hk}
	\affiliation{Department of Physics, The Chinese University of Hong Kong, Shatin, Hong Kong SAR, China}
	\affiliation{State Key Laboratory of Quantum Information Technologies and Materials, The Chinese University of Hong Kong, Shatin, Hong Kong SAR, China}
	
	\begin{abstract}
		Gauge fields are fundamental to modern physics, but prescribed gauge configurations are often difficult to implement in artificial systems. Here, we present a general scheme for realizing arbitrary static $\mathrm{O}(N)$ lattice gauge configurations using symmetry-protected zero modes of sublattice-imbalanced bipartite units. The target $\mathrm{O}(N)$ link on each bond is encoded in the connectivity and strengths of positive microscopic couplings. By decoupling the zero-mode manifold from the remaining modes, the target gauge Hamiltonian forms an exact spectral block of the microscopic tight-binding model rather than a perturbative approximation. We experimentally demonstrate this framework in acoustic crystals through a $\mathbb{Z}_2$ quadrupole topological insulator, an $\mathrm{SO}(2)$ Hofstadter model, and an $\mathrm{SO}(3)$ non-Abelian topological insulator. Our results provide a general and accessible route to gauge-field physics in artificial systems.
	\end{abstract}
	\maketitle
	
	\textit{Introduction.}---
	The concept of gauge fields, a mathematical framework that provides a local description of gauge-invariant physical laws, has served as a cornerstone in a broad range of areas in physics. Celebrated examples include the Yang–Mills framework underlying the Standard Model~\cite{YangMills1954} and the Aharonov-Bohm effect uncovering the physical reality of electromagnetic potentials~\cite{ABeffect1959}. In crystalline materials, gauge fields implemented through fluxes are intrinsically linked to topological phenomena, as exemplified by the Haldane model with nonzero local magnetic fluxes~\cite{Haldane1988PRL}, and by topological insulators (TIs) where the spin-orbit coupling serves as a static $\mathrm{SU}(2)$ gauge field~\cite{Kane2005PRL,Bernevig2006PRL}. More recently, lattice gauge fields have been shown to generate projective crystal symmetry algebras,  leading to unconventional effects such as M\"{o}bius boundary modes and Klein-bottle Brillouin zones~\cite{Zhao2020PRB,chen2022nc}.

	These theoretical advances have stimulated strong experimental interest in realizing gauge fields in artificial platforms~\cite{yang2024non}. In photonic and acoustic systems, $\mathbb{Z}_2$ gauge fields have been implemented by engineering effective negative couplings~\cite{serra2018observation,peterson2018nature,xue2020nc}. Real-space non-Abelian gauge fields have also been synthesized using nonreciprocal optical elements, synthetic dimensions, and circuit implementations~\cite{Yiyang2019science,wujie2022ne,cheng2025nature}. Nevertheless, implementing general matrix-valued gauge links remains challenging, particularly when their values must be tuned independently from bond to bond. Related enlarged-lattice approaches have recently encoded Abelian and non-Abelian gauge structures into representation sectors of Cayley--Schreier lattices with real hoppings~\cite{marciani2025translation,guba2025topological}. These constructions focus on gauge links associated with discrete groups, and different representation sectors may overlap in energy,so that the target gauge sector is not automatically spectrally isolated. A different line of work uses subsystem reductions to reveal latent symmetries or effective non-Hermitian topology in Hermitian systems~\cite{rontgen2021latent,hamanaka2024non}. There, coupling to the eliminated degrees of freedom generally produces an energy-dependent self-energy in the reduced Hamiltonian. We thus address a distinct gauge-engineering problem: how to implement independently programmable, generally continuous gauge links using only positive couplings, while spectrally isolating the resulting exact gauge sector.

	We solve this problem using symmetry-protected zero modes of sublattice-imbalanced bipartite units. Their sign-changing wave-function components generate arbitrary real effective hoppings from positive microscopic connections, enabling bond-by-bond implementation of prescribed $\mathrm{O}(N)$ gauge links. The couplings can also be chosen to eliminate matrix elements between the zero-mode manifold and its complement. Therefore, the projected Hamiltonian is an invariant block of the microscopic Hamiltonian, and a sufficiently large intra-unit gap separates it from the other bands. While the simplest case of $N=1$ corresponds to the $\mathbb{Z}_2$ gauge field and $\mathrm{SO}(2)$ is equivalent to $\mathrm{U}(1)$, our scheme applies in principle to all compact gauge groups, as any compact Lie group can be regarded as a subgroup of $\mathrm{O}(N)$ for sufficiently large $N$. We experimentally demonstrate this construction in acoustic crystals through a $\mathbb{Z}_2$ quadrupole topological insulator, an $\mathrm{SO}(2)$ Hofstadter model, and an $\mathrm{SO}(3)$ non-Abelian topological insulator. In particular, the $\mathrm{SO}(2)$ realization possesses continuously parameterizable link matrices, going beyond gauge connections restricted to discrete-group elements. Our results are fully based on simple models with only positive and short-range couplings, and are hence applicable to a broad range of artificial simulators, including photonic/acoustic crystals~\cite{lu2014np,Ozawa2019RMP,xue2022NRM}, mechanical lattices~\cite{huber2016NP,ma2019NRP}, electric circuits~\cite{sahin2025topolectrical}, and cold atoms~\cite{zhang2018adip,Cooper2019RMP}.
	
\begin{figure}[t]
		\centering
		\includegraphics[width=\columnwidth]{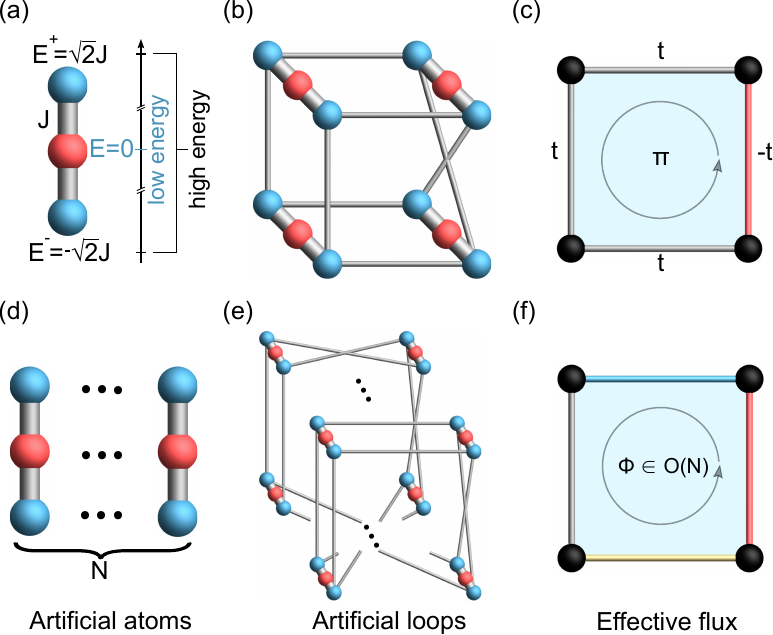}
		\caption{General framework. (a) The three-site unit comprising one $A$-site (red) and two $B$-sites (blue). This system possesses two high-energy modes $\pm \sqrt{2}J$ and one zero mode. (b) A plaquette composed of three-site units. (c) The corresponding effective model with $\pi$ flux, where the gray (red) bonds denote positive (negative) couplings, respectively. (d) The $\mathrm{O}(N)$ artificial atom consisting of $N$ copies of the three-site unit. (e) A plaquette constructed from the artificial atom in (d). (f) The corresponding effective model with $\mathrm{O}(N)$ flux.}\label{FIG1}
	\end{figure}

	\textit{General framework.}---
	The general principle of our construction leverages the well-known property of finite bipartite tight-binding models (also known as Lieb's theorem): if the numbers of $A$-sites and $B$-sites differ, the system hosts $n$ intrinsic symmetry-protected zero modes, where $n$ is the difference in their numbers~\cite{Lieb1989PRL,Brouwer2002PRB,Lieb_theorem}. Based on Lieb's theorem, we construct a basic unit with $M_A$ $A$-sites and $M_B$ $B$-sites, which contains $n=M_B-M_A$ zero modes (we assume $M_B>M_A$; see Fig.~\ref{FIG1}(a) for the simplest three-site example). An ``artificial atom" can then be built from identical copies of this unit. For generality and experimental feasibility, we further assume that all couplings are positive real numbers. To implement a gauge field, we restrict our discussion to the subspace spanned by the zero modes. This subspace can be safely decoupled from other modes by selecting appropriate inter-atom couplings. Then, a sufficiently large intra-unit gap can separate the zero-mode subspace from the other bands.
	
	The Hamiltonian of the artificial atom with $N$ basic units can be expressed in the block anti-diagonal form
	\begin{equation}\label{eq:antidiagonal} 
		\mathcal{H}_{\text{one-atom}} = \begin{bmatrix} 
			0 & \mathcal{Q} \\ 
			\mathcal{Q}^T & 0 
		\end{bmatrix}, 
	\end{equation} 
	where $\mathcal{Q}$ is a real matrix of dimension $NM_A \times NM_B$ with non-negative entries, and $\mathcal{Q}^T$ is its transpose. Hereafter, we take $n=M_B-M_A=1$ (i.e., one zero mode per basic unit) for simplicity, whereas the discussion can be easily generalized to cases with $n>1$. 
	The spectrum of $\mathcal{H}_{\text{one-atom}}$ consists of $NM_A$ pairs of nonzero eigenvalues $\{\pm \lambda_i\}$ and $N$ zero-energy modes
	\begin{equation}
		\{\lambda_1,-\lambda_1, \dots, \lambda_{NM_A},-\lambda_{NM_A},0,\dots, 0\}.
	\end{equation}
	The zero-mode eigenstates $|\psi_j\rangle$ with $j \in \{1, 2, \dots, N\}$ reside exclusively on the $B$-sites, and take the form
	\begin{equation}\label{eq:eigenHs}
		|\psi_j\rangle = \begin{pmatrix}
			\mathbf{0}_{NM_A} \\
			\mathbf{u}_{NM_A+j}
		\end{pmatrix}.
	\end{equation}
	Here, $\mathbf{0}_{NM_A}$ is a zero vector of length $NM_A$, and $\mathbf{u}_{k}$ denotes the $k$-th column vector of the orthogonal matrix $\mathcal{U}$ from the singular-value decomposition $\mathcal{Q}=\mathcal{V}\Sigma\mathcal{U}^T$, where $\mathcal{V}$ and $\mathcal{U}$ are orthogonal matrices of dimensions $NM_A$ and $NM_B$, and $\Sigma=[\Lambda,\mathbf{0}_{NM_A\times N}]$ with diagonal singular values $\Lambda$ ($\lambda_i \geq 0$). Since $\mathcal{Q}$ has rank $NM_A$, the last $N$ columns of $\mathcal{U}$ (indices $NM_A+1$ to $NM_A+N$) span the null space of $\mathcal{Q}$, thereby forming the basis for the zero-energy subspace.
	
	To characterize the effective coupling between artificial atoms, we consider a two-atom system with the microscopic Hamiltonian:
	\begin{equation}
		\H_{\text{two-atom}}=\begin{bmatrix}
			\H_{\text{one-atom}} & \Delta \\ \Delta^T & \H_{\text{one-atom}} 
		\end{bmatrix},
	\end{equation}
	where $\Delta$ represents the inter-atom hopping matrix. The zero-mode manifold of the decoupled atoms is spanned by the eigenvectors of
	$\sigma_0\otimes\H_{\text{one-atom}}$, i.e.,
	$|1\rangle\otimes|\psi_j\rangle$ and $|2\rangle\otimes|\psi_j\rangle$, where $|1\rangle$ and
	$|2\rangle$ label the two artificial atoms. Projecting onto the zero-mode manifold gives the zero-mode-sector effective Hamiltonian $\H_\text{eff}$ as
	\begin{equation}\label{eq:HEFFLA}
		\H_\text{eff}=\begin{bmatrix}
			0 & \Delta_\text{eff} \\
			\Delta^T_\text{eff} & 0
		\end{bmatrix}.
	\end{equation}
	The $N\times N$ effective hopping matrix $\Delta_\text{eff}$ is given by 
	\begin{equation}\label{eq:effDel}
		[\Delta_\text{eff}]_{jk}=\langle \psi_j|\Delta|\psi_k\rangle.
	\end{equation}
	In our construction, the inter-atom couplings are chosen such that $\langle\pm\lambda_i|\Delta|\psi_j\rangle=\langle\pm\lambda_i|\Delta^T|\psi_j\rangle=0$ for all $i$, $j$~(see Supplemental Material (SM) Sec.~I~\cite{SM}). Consequently, the projected Hamiltonian $\H_{\text{eff}}$ would be an exact block of the microscopic Hamiltonian rather than a perturbative low-energy approximation.
	
	Since the zero modes $|\psi_j\rangle$ may contain negative components, the effective couplings $[\Delta_{\text{eff}}]_{jk}$ can also be negative, despite all original couplings being positive. For the elementary example in Fig.~\ref{FIG1}(a), the system possesses two high-energy states $\pm \sqrt{2}J$ and one zero mode $|\psi\rangle = (0, 1, -1)^T / \sqrt{2}$ in the basis $(A, B, B)$. When two artificial atoms are connected through a parallel (cross) coupling, the effective coupling can be derived as $\Delta^p_{\text{eff}} = \langle \psi | \Delta^p | \psi \rangle = t$ ($\Delta^c_{\text{eff}} = \langle \psi | \Delta^c | \psi \rangle = -t$). This demonstrates that any $\mathbb{Z}_2$ gauge configuration can be readily realized by appropriately engineering the parallel and crossing couplings [Figs.~\ref{FIG1}(b) and \ref{FIG1}(c)]. 
	
	To realize the $\mathrm{O}(N)$ gauge fields, we can construct a $3N$-site artificial atom consisting of $N$ three-site units [Fig.~\ref{FIG1}(d)]. Since intra-atom couplings between units are zero, the zero modes are localized explicitly on the $B$-sites of the $j$-th unit:
	\begin{equation} 
		|\psi_j\rangle = \frac{1}{\sqrt{2}}(\mathbf{0}_{N+2j-2}, 1, -1, \mathbf{0}_{2N-2j})^T.
	\end{equation} 
	When coupling two such atoms, the effective hopping amplitudes within the zero-mode subspace are derived as:
	\begin{equation}\label{eq:O(N)eff} 
		[\Delta_{\text{eff}}]_{jk}=\frac{1}{2} (\Delta_{a,b} + \Delta_{a+1,b+1} - \Delta_{a,b+1} - \Delta_{a+1,b}), 
	\end{equation} 
	where indices $a = N + 2j - 1$ and $b = N + 2k - 1$ denote the relevant sites. Here, $[\Delta_{\text{eff}}]_{jk}$ represents the effective coupling between the $j$-th unit of the first atom and the $k$-th unit of the second. Similar to the previous case, Eq.~\eqref{eq:O(N)eff} demonstrates that parallel and crossing couplings contribute positively and negatively, respectively. It is a mathematical fact that any real effective hopping matrix is accessible, because the competition between parallel and crossing hoppings allows each element to take any real value. Thus, by appropriately engineering the coupling patterns, we can realize any configuration of the $\mathrm{O}(N)$ gauge field [Figs.~\ref{FIG1}(e) and \ref{FIG1}(f)]. 
	
\begin{figure}[t]
		\centering
		\includegraphics[width=\columnwidth]{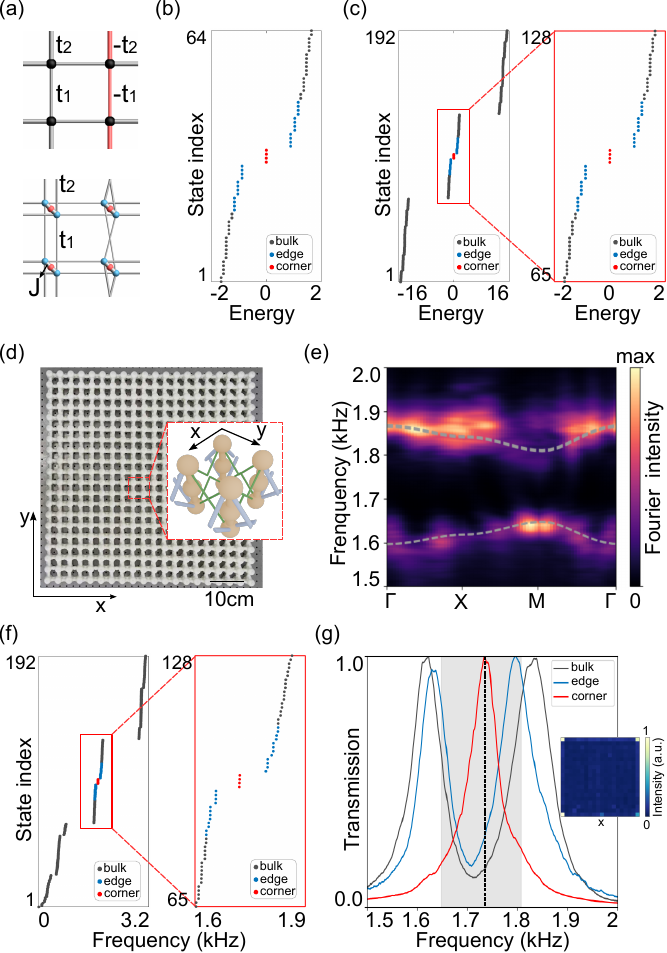}
		\caption{A quadrupole topological insulator with $\mathbb{Z}_2$ gauge field. (a) The target lattice model (top) and the constructed lattice model using $N=1$ artificial atoms (bottom). Gray (red) bonds represent positive (negative) couplings. (b) Calculated energy spectrum of the target model under a square geometry with $t_1=0.2$ and $t_2=1$. (c) Calculated energy spectrum of the constructed model under a square geometry with $J=10$, $t_1=0.2$ and $t_2=1$. (d) Photograph of the fabricated acoustic sample. The inset shows the unit cell. (e) Measured (color map) and simulated (black curves) bulk dispersion of the acoustic crystal. (f) Simulated eigenfrequencies of the acoustic crystal under a square geometry. (g) Measured transmission spectra at bulk (gray), edge (blue), and corner (red) regions. The inset displays the acoustic intensity distribution at 1735 Hz (indicated by the dashed line).
		} \label{FIG2}
	\end{figure}

	\textit{A quadrupole topological insulator with $\mathbb{Z}_2$ gauge field.}---
	We first apply our framework to a quadrupole TI with a $\pi$ flux through each plaquette [Fig.~\ref{FIG2}(a), top]~\cite{SM, Wladimir2017Science}. The flux enforces anticommuting reflection symmetries $\textsf{M}_x$ and $\textsf{M}_y$, quantizing the bulk quadrupole moment and producing four midgap corner states in a finite lattice [Fig.~\ref{FIG2}(b)]. We replace each target-lattice site with an $N=1$ artificial atom and each positive (negative) hopping with a parallel (crossing) coupling. The resulting lattice contains only positive microscopic couplings [Fig.~\ref{FIG2}(a), bottom] and preserves modified reflections $\textsf{M}'_{x,y}$, whose projections onto the zero-mode subspace also anticommute~\cite{SM}. Its zero-mode spectrum therefore reproduces that of the target model, including the four corner states [Fig.~\ref{FIG2}(c)].

	We implement this model using coupled acoustic resonators [Fig.~\ref{FIG2}(d)]~\cite{xue2022NRM, xiao2015synthetic, xue2019acoustic}. Here, each site is realized by a spherical resonator and the couplings are mediated through cylindrical tubes (see SM Sec.~X for structural details~\cite{SM}). The whole structure is filled with air and surrounded by rigid walls. By tuning the coupling tubes' radii, we open a bandgap around 1735 Hz, within which four corner states appear [Figs.~\ref{FIG2}(e) and \ref{FIG2}(f)]. To characterize the sample, we experimentally measure the transmission spectra and intensity distributions~\cite{SM}. As plotted in Fig.~\ref{FIG2}(g), the transmission exhibits peaks at the predicted eigenfrequencies. At the corner state frequency, the measured field is strongly localized at the corners [see the inset of Fig.~\ref{FIG2}(g)].
	
	\begin{figure*}[t]
		\centering
		\includegraphics[width=\textwidth]{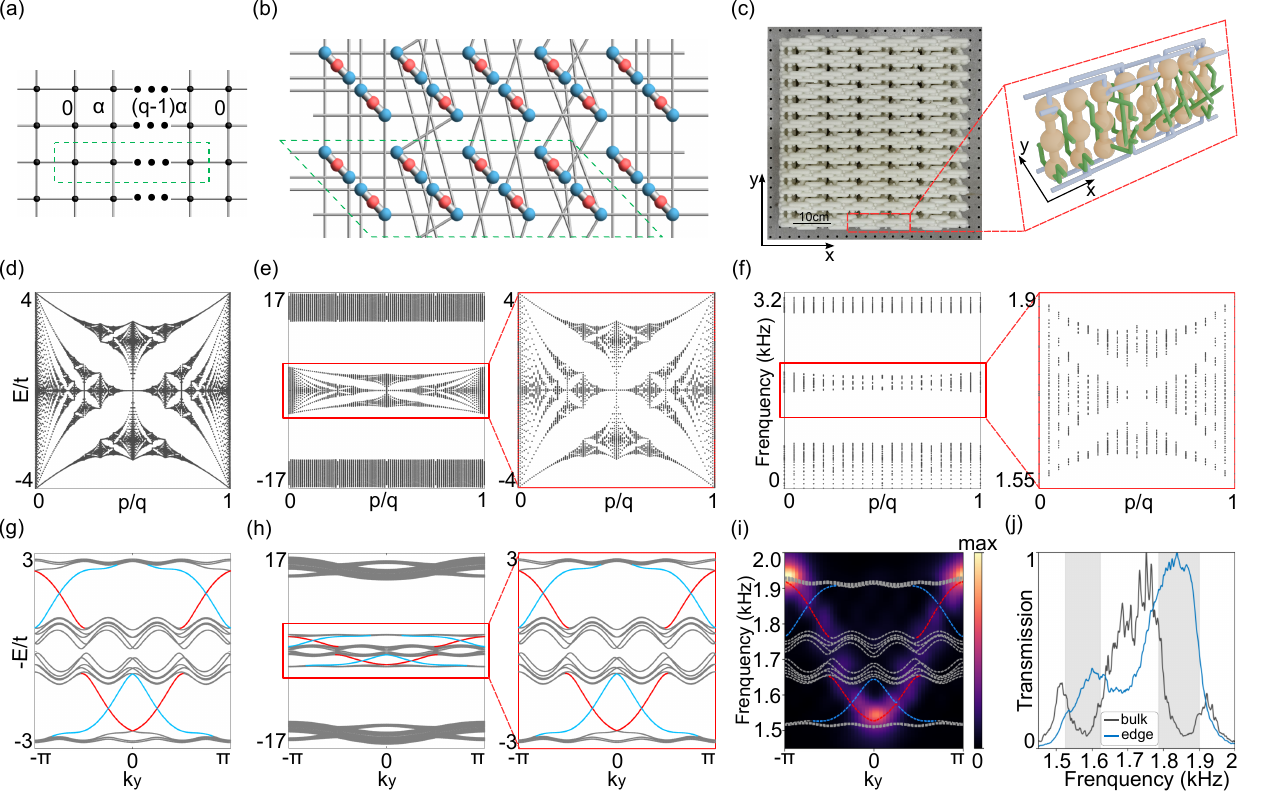}
		\caption{An $\mathrm{SO}(2)$ Hofstadter model. (a) The target $\mathrm{SO}(2)$ Hofstadter lattice model, with hopping phases indicated. 
			(b) The constructed lattice model  with $\alpha=\pi/2$. Dashed lines mark the unit cell. 
			(c) Photograph of the fabricated acoustic sample with three and twelve unit cells along the $x$ and $y$ directions, respectively. The right panel illustrates the unit cell. 
			(d) The Hofstadter butterfly spectrum calculated from the target lattice model, plotted against the flux ratio $p/q$ (fixed $q=100$). 
			(e) Energy spectrum of the constructed lattice model, where the Hofstadter butterfly emerges in the low-energy window (red box). 
			(f) Simulated eigenfrequencies as a function of $p/q$ for acoustic crystals with fixed $q=20$.
			(g) Calculated edge spectrum of the target model in (a)  under open boundary conditions along the $x$ direction. The red (blue) curves correspond to left (right) edge states. 
			(h) Calculated edge spectrum of the constructed model in (b) under open boundary conditions along the $x$ direction.
			(i) Measured (color map) and simulated (dashed curves) dispersions on the left boundary of the acoustic crystal. 
			(j) Measured bulk (gray) and edge (blue) transmissions of the acoustic crystal.
		} \label{FIG3}
	\end{figure*}
	
	\textit{An $\mathrm{SO}(2)$ Hofstadter model.}---
	In the Hofstadter model~\cite{hofstadter1976energy}, each plaquette carries a $\mathrm{U}(1)$ gauge flux, which is challenging to realize due to the broken time-reversal symmetry. Since $\mathrm{U}(1) \cong \mathrm{SO}(2)$, we instead consider an $\mathrm{SO}(2)$ Hofstadter model by transforming the complex hopping term $te^{i\phi}$ into the real hopping matrix $te^{i\phi\sigma_2}$. The corresponding lattice model is depicted in Fig.~\ref{FIG3}(a), with all hoppings along the $x$ ($y$) direction given by $t\sigma_0$ ($te^{im\alpha\sigma_2}$). Here, $m$ labels the sites along the $x$-direction, and $\alpha = 2\pi p/q$ with $p, q \in \mathbb{N}^+$. Consequently, each plaquette carries an $\mathrm{SO}(2)$ gauge flux, $\Phi = e^{-i(m-1)\alpha\sigma_2} e^{im\alpha\sigma_2} = e^{i\alpha\sigma_2}$.

	To realize this model, we use $N=2$ artificial atoms, with positive and negative effective couplings again achieved by parallel and crossing couplings, respectively. A specific lattice structure for $\alpha = \pi/2$ and the corresponding acoustic design are illustrated in Fig.~\ref{FIG3}(b) and \ref{FIG3}(c), respectively~\cite{SM}. The target model exhibits the Hofstadter butterfly as a function of $p/q$ [Fig.~\ref{FIG3}(d)], which is reproduced exactly in the low-energy sector of the constructed model [Fig.~\ref{FIG3}(e)]. Acoustic simulations with fixed $q = 20$ recover the same fractal spectrum [Fig.~\ref{FIG3}(f)].
	
	The nontrivial bulk topology associated with the $\mathrm{SO}(2)$ flux produces gapless edge states under open boundary conditions [Figs.~\ref{FIG3}(g) and (h)]~\cite{SM,minus_sign}. These states are observed experimentally in the Fourier-transformed edge field [Fig.~\ref{FIG3}(i)] and as peaks in the edge transmission spectrum [Fig.~\ref{FIG3}(j)].
	
	\begin{figure*}[t]
		\centering
		\includegraphics[width=\textwidth]{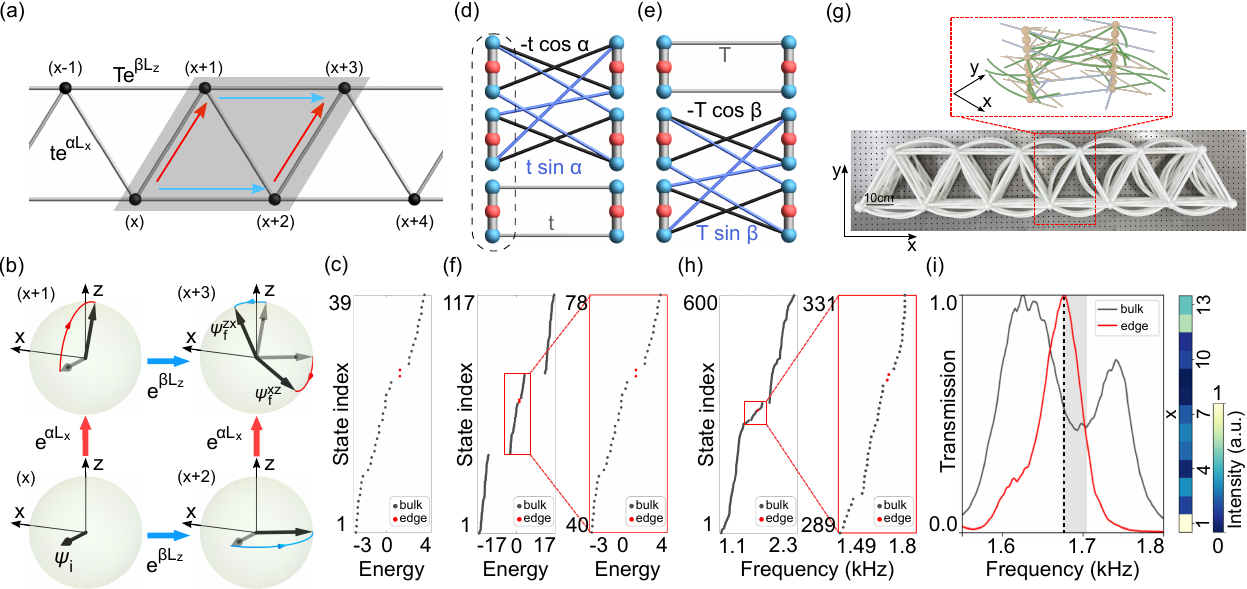}
		\caption{A non-Abelian $\mathrm{SO}(3)$ gauge field. 
			(a) The target lattice model featuring an $\mathrm{SO}(3)$ gauge field. Each site $(x)$ hosts three orbitals, connected by the indicated hopping matrices. 
			(b) Illustration of the non-Abelian nature of the $\mathrm{SO}(3)$ lattice gauge field. A particle initialized in state $\psi_i$ at site $x$ evolves into distinct final states when transported to site $x+3$ along two different pathways. 
			(c) Calculated energy spectrum of the target model in (a) under open boundary conditions. 
			(d) and (e) Physical realizations of the hopping terms $te^{\alpha L_x}$ and $Te^{\beta L_z}$ utilizing $N=3$ artificial atoms. 
			(f) Calculated energy spectrum of the constructed model. 
			(g) Photograph of the acoustic sample. The top panel details the unit cell structure. 
			(h) Simulated eigenfrequencies of the acoustic crystal under open boundary conditions. 
			(i) Measured transmission spectra (left panel) and acoustic intensity distributions at the frequency $1672$ Hz indicated by the black dashed line (right panel).} \label{FIG4}
	\end{figure*}
	
	\textit{A non-Abelian $\mathrm{SO}(3)$ gauge field.}---
	Finally, we demonstrate a non-Abelian $\mathrm{SO}(3)$ gauge field on a 1D lattice model (see SM Sec.~IX for a 2D model~\cite{SM}). Each unit cell consists of a single site with three orbitals, with oblique and horizontal hoppings $te^{\alpha L_x}$ and $Te^{\beta L_z}$, respectively, where $L_{i}$ are the real antisymmetric generators  of the Lie algebra $\mathfrak{so}(3)$~\cite{SM} [Fig.~\ref{FIG4}(a)]. The gauge field is non-Abelian because transport depends on the order of the link operations. For the two paths from $(x)$ to $(x+3)$ shown in Fig.~\ref{FIG4}(a), an initial state $\psi_i$ evolves into $\psi_f^{zx}=e^{\beta L_z}e^{\alpha L_x}\psi_i$ and $\psi_f^{xz}=e^{\alpha L_x}e^{\beta L_z}\psi_i$, respectively [Fig.~\ref{FIG4}(b)]. These final states generally differ because $[L_x,L_z]\ne 0$.
	
	The model respects $\mathcal{PT}$ symmetry, where $\hat{\mathcal{P}}=e^{\pi L_y}\I=\text{diag}\{-1, 1, -1\}\I$ is a gauge-dressed inversion symmetry~\cite{Switch_PT} and $\hat{\mathcal{T}}=\K\I$ is the time-reversal symmetry. Here, $\K$ denotes the complex conjugation and $\I$ denotes the inversion of momenta. For a $\mathcal{PT}$-symmetric non-degenerate $3\times 3$ Hamiltonian, the classifying space is $M_3 = \mathrm{O}(3)/(\mathrm{O}(1) \times \mathrm{O}(1) \times \mathrm{O}(1))$, with the fundamental group being the non-Abelian quaternion group $\{\pm1,\pm \text{i},\pm \text{j},\pm \text{k}\}$~\cite{Wuscience2019}. For $t=-1.2$, $T=-1$, $\alpha=5\pi/8$, and $\beta = 2\pi/3$, the system has charge $\text{i}$ and supports in-gap edge states [Fig.~\ref{FIG4}(c)]~\cite{SM,minus_sign}.
	
	We implement the model with $\alpha=5\pi/8$ and $\beta = 2\pi/3$ using $N=3$ artificial atoms, replacing the oblique and horizontal couplings with the hopping configurations detailed in Figs.~\ref{FIG4}(d) and \ref{FIG4}(e), respectively~\cite{SM}. For $J=-10$, the constructed model reproduces the target spectrum exactly within its zero-mode sector [Fig.~\ref{FIG4}(f)]~\cite{minus_sign}. The corresponding acoustic crystal contains 13 unit cells [Fig.~\ref{FIG4}(g)]. Its simulated spectrum exhibits the predicted bulk and edge states [Fig.~\ref{FIG4}(h)], while the measured transmission peaks agree with their simulated frequency ranges [Fig.~\ref{FIG4}(i)]. The field distribution at 1672Hz further confirms localization at the boundary [Fig.~\ref{FIG4}(i)].

	\textit{Discussion.}---
	We have established a general zero-mode construction for realizing prescribed static $\mathrm{O}(N)$ lattice gauge configurations. In contrast to enlarged-lattice constructions developed for discrete gauge groups~\cite{marciani2025translation,guba2025topological}, our approach directly programs arbitrary, generally continuous $\mathrm{O}(N)$ link matrices bond by bond and spectrally isolates the resulting exact gauge sector. The three acoustic realizations illustrate the same design principle for discrete Abelian, continuous Abelian, and non-Abelian links. Since only positive couplings are involved, and $N$ and $n$ can be further increased, our approach allows for the construction of complicated models with a high number of orbitals and even in three dimensions~\cite{SM}, and is readily realizable in other artificial crystals~\cite{lu2014np,Ozawa2019RMP,xue2022NRM,huber2016NP,ma2019NRP,sahin2025topolectrical}.
	
	\begin{acknowledgments}
		\textit{Acknowledgments.}---
		This work was supported by the National Natural Science Foundation of China under Grant No.~62401491 (H.X.), the National Key Research and Development Program of China under Grant No.~2025YFA1412300 (H.X.), the Research Grants Council of the Hong Kong SAR, under Grant Nos.~24304825 (H.X.), 17301224 (Y.X.Z.) and 17302525 (Y.X.Z.), and the Guangdong Provincial Quantum Science Strategic Initiative under Grant Nos.~GDZX2501012 (H.X.) and GDZX2504003 (Y.X.Z.).
	\end{acknowledgments}
	
	\bibliography{references}

\begin{thebibliography}{37}
\bibitem [{Lie()}]{Lieb_theorem}%
  \BibitemOpen
  \href@noop {} {}\bibinfo {note} {A bipartite system is characterized by a
  Hamiltonian $\H$ that couples two distinct sublattices, $A$ and $B$, with
  $M_A$ and $M_B$ sites, respectively. Such a system possesses a sublattice
  (chiral) symmetry operator $\mathcal{S}$, which anticommutes with the
  Hamiltonian, namely, $\{\H, \mathcal{S}\} = 0$. This symmetry dictates that
  the energy spectrum is symmetric around zero: for every eigenstate
  $|\psi_E\rangle$ with energy $E$, there exists a partner state
  $\mathcal{S}|\psi_E\rangle$ with energy $-E$. To form such a pair, we need
  one degree of freedom from $A$ and one degree of freedom from $B$ to couple
  together. If the sublattice sizes are unequal (e.g., $M_B > M_A$), the
  pairing of nonzero energy states implies that the number of non-zero modes
  cannot exceed $2M_A$. Consequently, the remaining $n = M_B - M_A$ states are
  forced to reside at zero energy ($E=0$). These zero modes are intrinsic to
  the lattice topology and are protected by the sublattice
  symmetry.}\BibitemShut {Stop}%
\bibitem [{min()}]{minus_sign}%
  \BibitemOpen
  \href@noop {} {}\bibinfo {note} {To facilitate direct comparison with the
  acoustic experiment, Figs.~3(g) and (h) plot $-E/t$. In the $\mathrm{SO}(3)$
  example, the tight-binding parameters are written with an overall negative
  sign to match the spectral convention used for the acoustic realization.
  These sign conventions reverse the spectral ordering but leave the eigenmodes
  and their boundary localization unchanged.}\BibitemShut {Stop}%
\end{thebibliography}
	
\end{document}